\pdfoutput=1
\documentclass[conference]{IEEEconf}

\input epsf
\usepackage{graphicx}
\usepackage{multirow}
\usepackage{titlesec}
\usepackage{balance}
\usepackage{stfloats}
\usepackage{amssymb}
\usepackage{booktabs}
\usepackage{hyperref}
\usepackage[linesnumbered,ruled,vlined]{algorithm2e}
\usepackage{amsthm}
\theoremstyle{definition}
\newtheorem{pcrule}{Rule}
\usepackage{threeparttable}
\usepackage{enumitem}
\usepackage{listings,listings-rust}
\usepackage[svgnames]{xcolor}

\newcommand{\CellWithForcedBreak}[2][c]{%
\begin{tabular}[#1]{@{}c@{}}#2\end{tabular}}

\renewcommand{\thetable}{\arabic{table}}  

\renewcommand\thesection{\arabic{section}} 
\renewcommand\thesubsection{\thesection.\arabic{subsection}}
\renewcommand\thesubsubsection{\thesubsection.\arabic{subsubsection}}

\begin{document}

\title{\textbf{\Large PinChecker: Identifying Unsound Safe Abstractions of Rust Pinning APIs\\}}

\author{Yuxuan Dai, Yang Feng$^{*}$\\
	\normalsize State Key Laboratory for Novel Software Technology,
Nanjing University, Nanjing, 210023, China\\
	\normalsize yxdai@smail.nju.edu.cn, fengyang@nju.edu.cn\\
	\normalsize *corresponding author
}


\maketitle
\begin{abstract}
The pinning APIs of Rust language guarantee memory location stability for self-referential and asynchronous constructs, as long as used according to the pinning API contract. Rust ensures violations of such contract are impossible in regular safe code, but not in unsafe code where unsafe pinning APIs can be used. Library authors can encapsulate arbitrary unsafe code within regular library functions. These can be freely called in higher-level code without explicit warnings. Therefore, it is crucial to analyze library functions to rule out pinning API contract violations. Unfortunately, such testing relies on manual analysis by library authors, which is ineffective.
Our goal is to develop a methodology that, given a library, attempts to construct programs that intentionally breach the pinning API contract by chaining library function calls, thereby verifying their soundness. We introduce RPIL, a novel intermediate representation that models functions' critical behaviors pertaining to pinning APIs. We implement PinChecker, a synthesis-driven violation detection tool guided by RPIL, which automatically synthesizes bug-revealing programs. Our experiments on 13 popular Rust libraries from crates.io found 2 confirmed bugs.
\end{abstract}
\IEEEoverridecommandlockouts
\vspace{1.5ex}

%
\IEEEpeerreviewmaketitle

\section{Introduction}
Rust has gained widespread attention and adoption across multiple domains including operating systems, embedded devices, and high-performance web frameworks, thanks to its efficient compiler and strict type system. In practice, Rust strictly controls memory access through core mechanisms such as ownership and borrowing checks, combined with lifetime inference to avoid common issues like dangling pointers and data races, thereby enhancing program safety. However, this safety mechanism can restrict flexibility in scenarios that require low-level operations or custom memory management. To accommodate such demands, Rust allows the use of \texttt{unsafe} to bypass certain compiler safety checks. Moreover, Rust permits encapsulating unsafe code inside regular non-\texttt{unsafe} functions, a practice known as creating a \emph{safe abstraction}~\cite{RustPLUnsafeRust2025}. The encapsulating functions can be freely called in higher-level code without explicit warnings. An empirical study on open-source projects reveals that approximately 42.8\% of safety-related patches revolve around the \texttt{unsafe} keyword~\cite{BeyondMemorySafety2024}, indicating latent security risks in the use of \texttt{unsafe} and safe abstractions.

To help developers in using \texttt{unsafe} and creating safe abstractions correctly, the Rust community has compiled documents such as \emph{Rust's Unsafe Code Guidelines Reference}~\cite{RustUCG2025}, providing practical recommendations on how to avoid common errors. For instance, library developers are encouraged to document the safety conditions of their safe abstraction functions to guide proper usage~\cite{SafetyComments2025}, although many libraries still lack such non-mandatory declarations~\cite{UnsafeEncapsulation2024}. Additionally, both academia and industry have introduced various analysis tools \cite{JungPOPL2020,QinPLDI2020,XuMSChallenge2020} that identify potential safety issues. Nevertheless, most projects continue to rely heavily on manual inspection to verify edge cases related to \texttt{unsafe} code~\cite{HowUseUnsafe2020,IsUsedSafely2020}. In this context, we focus on a particular standard library feature known as the pinning API~\cite{PinStd2025,PinStruct2025}. We investigate whether safe abstractions built around this Pin mechanism in user-created Rust libraries might violate the pinning API contract~\cite{PinStruct2025}.

The Pin mechanism ensures memory location stability for self-referential structures through the type system, serving as a cornerstone for Rust's asynchronous programming implementation~\cite{PinStd2025}. When developers implement asynchronous tasks through the \texttt{Future} trait, the compiler constructs a state-machine data structure containing self-referential pointers, which must rely on \texttt{Pin}'s pinning constraint semantics to prevent self-referential pointer invalidation due to memory moves. Although the standard library documentation~\cite{PinStruct2025} provides warnings about avoiding improper encapsulation in each method's ``Safety'' section, there remains a lack of specialized automated detection tools for pinning API contract violations.

The challenge lies in that pinning API contract violations (henceforth referred to as \emph{pin violations}) often remain latent within library functions. The violations manifest only when users compose these functions into complete programs. This makes it necessary to generate test programs incorporating the library's functions to verify the soundness of safe abstractions for pinning APIs. While existing program generation tools such as JCrasher~\cite{JCrasher04}, Randoop~\cite{Randoop07}, and RESTler~\cite{RESTler19} are available, they primarily rely on random or feedback-driven mutation strategies and are designed specifically for languages like Java. These tools lack understanding of Rust's pinning APIs and their semantics; direct adaptation would result in the generation of test cases irrelevant to pinning API verification, rendering them ineffective for our purpose.

To address these limitations, we present a constraint solving based automated program synthesis tool specifically designed for Rust pinning APIs. Our approach first translates library functions' critical behaviors into a novel lightweight intermediate representation \emph{RPIL}, which precisely models operations affecting values' pinning states and reference relationships. A reversible interpreter implemented in Prolog then leverages RPIL to synthesize minimal programs that trigger pin violations through constraint solving.

We implemented the proposed techniques in PinChecker, an extensible, synthesis-driven violation detection tool for Rust libraries encapsulating unsafe pinning APIs. We evaluated PinChecker on 13 popular Rust libraries from crates.io and discovered 2 pin soundness bugs. Our experimental results demonstrate that PinChecker can effectively identify unsound safe abstractions of pinning APIs with reasonable time cost. Furthermore, our evaluation shows that the reversible interpreter design significantly improves detection efficiency compared to conventional forward-only approaches. We make the following key contributions:

\begin{itemize}[]

\item We introduce RPIL, a novel intermediate representation that models functions' critical behaviors pertaining to pinning APIs.
\item Based on RPIL, we implement a violation detection tool called PinChecker for Rust libraries encapsulating unsafe pinning APIs.
\item We evaluate PinChecker on popular Rust libraries and demonstrate that it can generate minimal pin violation programs that expose bugs.
\end{itemize}

\section{Background and Motivating Example}
\begin{figure}[!tbp]
  \centering
  \includegraphics[width=\linewidth]{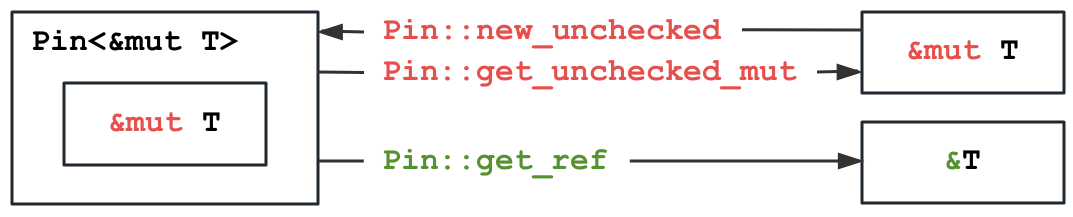}
  \caption{Basic usage of Rust pinning APIs.}
  \label{fig:pin-basic-usage}
\end{figure}

In Rust, the standard library's \texttt{pin} module provides a series of API functions for creating and manipulating Pin pointers. These functions are categorically divided into regular functions and unsafe functions. As shown in Figure~\ref{fig:pin-basic-usage}, function names in green represent ordinary APIs that can be invoked in non-unsafe contexts, such as \texttt{Pin::get\_ref} which extracts an immutable version (\texttt{\&T}) of the reference encapsulated by a Pin pointer. Since immutable references cannot trigger memory moves, there is no risk of pinning API contract violations. In contrast, functions names in red represent unsafe APIs, such as \texttt{Pin::new\_unchecked} for creating Pin pointers and \texttt{Pin::get\_unchecked\_mut} for extracting mutable references (\texttt{\&mut T}) from Pin pointers. These functions intentionally include ``\texttt{unchecked}'' in their names and must be called within unsafe contexts. Because they manipulate unrestricted mutable references, they can be directly used in memory move operations like \texttt{mem::swap}, which may introduce risks of pinning API contract violations~\cite{PinStd2025}.

The core principle of the pin contract is to encapsulate mutable references (\texttt{\&mut T}) so that only immutable references can be freely obtained, while mutable references can only be acquired in unsafe contexts through specific unsafe function calls. Consequently, any behavior that may violate the pin API contract can only occur within unsafe contexts, ensuring that the pinning API contract cannot be violated in ordinary contexts without unsafe code.


To demonstrate the definition of self-referential structures and related pin API contract violations, Figure~\ref{fig:self-ref-struct-code} presents an example of a self-referential struct, showing its definition and implementation. The struct contains two data members: one stores concrete data, while the other holds a pointer pointing to the former. The struct includes three operations: 1) Creation (\texttt{new}): creates an instance of the struct and assigns a concrete value to member \texttt{field1}; 2) Initialization (\texttt{init}): assigns a raw pointer to member \texttt{field2}, whose value is the address of the instance's member \texttt{field1}; 3) Validation (\texttt{validate}): verifies whether the address of member \texttt{field1} matches the address stored in member \texttt{field2}. The validation operation's method \texttt{SelfRef::validate} accepts a Pin pointer rather than an ordinary mutable reference as its parameter; consequently, this function must be called on a Pin pointer. Such design is common in practice to ensure that mutable references cannot be obtained within the function, thereby protecting the integrity of self-referential data.

\begin{figure}[!tbp]
  \begin{lstlisting}[language=Rust,style=colouredRust,basicstyle=\linespread{0.8}\footnotesize\ttfamily]
pub struct SelfRef {
  field1: i32,
  field2: Option<*const i32>,
  _marker: std::marker::PhantomPinned,
}

impl SelfRef {
  fn new() -> Self {
    Self {
      field1: 42,
      field2: None,
      _marker: std::marker::PhantomPinned,
    }
  }

  fn init(&mut self) {
    self.field2 = Some(&self.field1 as *const i32);
  }

  fn validate(self: Pin<&mut Self>) {
    let addr = &self.as_ref().field1 as *const i32;
    let stored_addr = self.as_ref().field2.unwrap();
    assert_eq!(addr, stored_addr);
  }
}
  \end{lstlisting}
  \caption{Example self-referential struct implementation.}
  \label{fig:self-ref-struct-code}
\end{figure}

To further illustrate the design of Pin APIs and how unsound safe abstractions can lead to pinning API contract violations, suppose we are developing a Rust library named \texttt{mylib}, with the above self-referential struct as part of it. The typical workflow is: creating an instance of the struct, calling \texttt{init} to establish a self-reference, then creating a Pin pointer to the struct and calling \texttt{validate} via that Pin pointer to verify the validity of the self-reference.

For user convenience in creating Pin pointers, we might define a \texttt{pin\_new} function that directly accepts a \texttt{\&mut T} to produce a \texttt{Pin<\&mut T>}, effectively providing a safe abstraction over the originally unsafe Pin function. Such safe abstraction is risky as it overlooks the fact that the pin contract's effectiveness relies on restricting the acquisition of mutable references to unsafe contexts only. Specifically, although this function constructs a Pin pointer, it doesn't prevent the acquisition of mutable references. As shown in the erroneous usage example in Figure~\ref{fig:pin-violation}, although Pin pointer \texttt{v2} is created, one can still obtain a regular \texttt{\&mut v1}, use it to perform memory-moving operations, and ultimately violate the pinning API contract.

\begin{figure}[!tbp]
\begin{minipage}[t]{\textwidth}
\textbf{A safe abstraction function introduced in \texttt{mylib}:}
\begin{lstlisting}[language=Rust,style=colouredRust,basicstyle=\linespread{0.8}\footnotesize\ttfamily]
pub fn pin_new<T>(r: &mut T) -> Pin<&mut T> {
  unsafe { Pin::new_unchecked(r) }
}
\end{lstlisting}
\end{minipage}
\hfill
\begin{minipage}[t]{\textwidth}
\vspace{0.64em}
\textbf{Erroneous usage example:}
\begin{lstlisting}[language=Rust,style=colouredRust,basicstyle=\linespread{0.8}\footnotesize\ttfamily]
let mut v1 = SelfRef::new();
v1.init();
<@\textcolor{FireBrick}{\textbf{let} v2 = mylib::pin\_new(\&\textbf{mut} v1);}@>
v2.validate();
<@\textcolor{FireBrick}{mem::swap(\&\textbf{mut} v1, ...);}@>
\end{lstlisting}
\end{minipage}
\hfill
\begin{minipage}[t]{\textwidth}
\vspace{0.64em}
\textbf{Expected usage example:}
\begin{lstlisting}[language=Rust,style=colouredRust,basicstyle=\linespread{0.8}\footnotesize\ttfamily]
let mut v1 = SelfRef::new();
v1.init();
<@\textcolor{ForestGreen}{\textbf{let} v1 = mylib::pin\_new(\&\textbf{mut} v1);}@>
v1.validate();
\end{lstlisting}
\end{minipage}
\caption{An unsound safe abstraction function and its usage.}
\label{fig:pin-violation}
\end{figure}

One possible correct usage is shown in the expected usage example in Figure~\ref{fig:pin-violation}. By shadowing the original variable which employs variable shadowing to override the original variable, the new Pin pointer becomes the sole medium through which one accesses the underlying value, thereby preventing the possibility of pinning API contract violations. However, as the above analysis shows, \texttt{mylib}'s \texttt{pin\_new} safe abstraction function conceals a security flaw: its correctness depends on callers following a strict and error-prone usage pattern. Once the caller fails to shadow the original variable, a direct mutable reference can still be acquired, which may enable memory move operations that threat pin soundness.

Hence, if a library function provides a seemingly safe but actually unsound abstraction of an unsafe Pin operation, it can enable constructing programs that violate pinning API contract, leading to severe safety vulnerabilities. This is precisely the risk our work aims to identify and address.

\section{Constraint-Based Detection of Pin Violation}
According to the Rust documentation, pinning contract violations manifest in two ways: (1) moving values that have already been pinned, and (2) failing to guarantee that pinned objects will eventually be dropped (the requirement that pinned objects must be dropped is also referred to as the \emph{Drop Guarantee}~\cite{PinStd2025}).

\begin{figure*}[!tbp]
  \centering
  \includegraphics[width=1.0\linewidth]{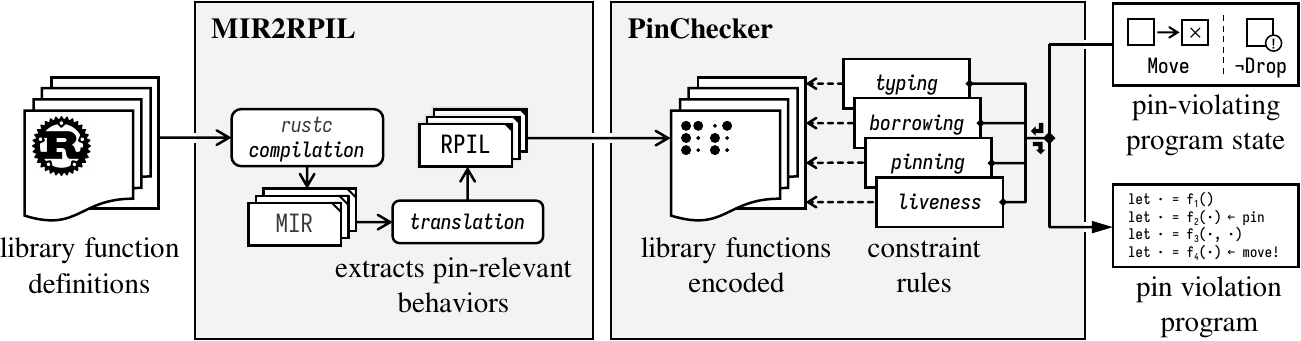}
  \caption{The workflow of PinChecker's components.}
  \label{fig:system-diagram}
\end{figure*}

In this study, we propose a constraint solving based approach for detecting pin API contract violations in Rust libraries. This approach automatically synthesizes minimal counterexample programs that expose unsound safe abstractions, thereby helping developers to uncover vulnerabilities that may lead to pin violations. As illustrated in Figure~\ref{fig:system-diagram}, the overall process is as follows: Given the \texttt{Cargo.toml} file path of a target Rust library, the \emph{MIR2RPIL} component first extracts the RPIL representations of all public functions in the library. It then hands these representations to the \emph{PinChecker} detection tool, which attempts to construct counterexample Rust programs that cause a pin violation, and finally returns them back to the user for verification. The core idea of this process is to abstract away the parts of the Rust program that are directly related to the pinning API contract, thereby reducing analysis complexity and enabling detection of pin violations through automated means.

\textbf{On Program Structure.} We abstracted the Rust programs by transforming them into linear sequences of single-line function calls. Specifically, each line of code calls a single function and assigns its result to a new variable. Figure~\ref{fig:rpil-and-context} shows the \emph{linear single-line function call form} corresponding to the erroneous usage example shown in Figure~\ref{fig:pin-violation}. This linearized format makes it easier to split the program into multiple steps and track the evolution of variables and the program context at each step, facilitating subsequent formal analysis.

\begin{figure}[!tbp]
  \centering
  \begin{lstlisting}[language=Rust,style=colouredRust,basicstyle=\linespread{0.8}\scriptsize\ttfamily]
let v1 = SelfRef::new();        // ;
let v2 = borrow_mut(v1);        // BORROW(v2, v1);
let v3 = SelfRef::init(v2);     // BORROW(v2[2][1], v2[1]);
let v4 = mylib::pin_new(v2);    // DEREF-PIN(v2);
let v5 = SelfRef::validate(v4); // ;
let v6 = deref_move(v2);        // DEREF-MOVE(v2);
-----------------------------------------------------------
line 1: { }
line 2: { v2->v1 }
line 3: { v2->v1, v2[2][1]->v2[1] }
line 4: { v2->v1, v2[2][1]->v2[1], v1:pinned }
line 5: { v2->v1, v2[2][1]->v2[1], v1:pinned }
line 6: { v2->v1, v2[2][1]->v2[1], v1:pinned_moved }
  \end{lstlisting}
  \caption{The linear form of a program and the RPIL instruction sequence corresponding to each function (above), the program context after executing the current line (below).}
  \label{fig:rpil-and-context}
\end{figure}

\textbf{On Aggregate Value Structure.} To precisely describe how composite types (such as structs, enums, and arrays) are used in programs, we introduced a sequential index based labeling scheme to uniquely number each substructure within an aggregate value. Specifically, struct fields are numbered according to their declaration order, enum variants and their internal fields are numbered similarly, and array elements are marked as different substructures based on their index positions. In this way, any deep access into an aggregate value can be represented with a notation like ``variableName[substructure index]'', facilitating subsequent formal analysis. For instance, consider the composite type definitions in Figure~\ref{fig:composite-type-definitions}, deep access expressions like \texttt{my\_var.unwrap().ptr} are represented in the form \texttt{my\_var[1][2]}.

\begin{figure}[!tbp]
  \centering
  \begin{lstlisting}[language=Rust,style=colouredRust,basicstyle=\linespread{0.8}\footnotesize\ttfamily]
struct SelfRef {
  data: i32,        // "data" labeled as [1]
  ptr: *const i32,  // "ptr" labeled as [2]
}

enum Option<T> {
  None,
  Some(T), // content of "Some" labeled as [1]
}

let my_var = Option::Some(SelfRef { ... } );
\end{lstlisting}
\caption{Example composite type definitions.}
  \label{fig:composite-type-definitions}
\end{figure}

\textbf{On Variable and Value States.} We employ the concept of program context, to maintain state descriptions for all variables and their substructures at each line of program execution. The context consists of three components $(\mathcal{V}, \mathcal{R}, \mathcal{S})$, where the variable set $\mathcal{V}$ records currently live variables and their type information, directed graph $\mathcal{R}$ represents the reference relationship graph, using directed edge $p\!\leadsto\!q$ to indicate value $p$ references value $q$, and the state space $\mathcal{S}$ tracks state changes each value (including substructures within composite types) undergoes during execution, such as whether it has been pinned or forgotten for destruction. For instance, a value in its initial state transitions to \texttt{pinned} state after being pinned. Further state transitions are described in detail in in Section~\ref{sec:pinchecker-violations}. This approach enables precise tracking of variable availability, value reference relationships, and value states required for pin violation checking.

\textbf{On Function Behavior.} We divide a function call into two components: static behavior and dynamic behavior. Static behavior is reflected by type inference information carried in function signatures, which determine the types of new variables created and added to variable set $\mathcal{V}$ after each function call. Dynamic behavior is captured by an intermediate representation called RPIL (Reference Provenance Intermediate Language), defined in this study. RPIL refines all operations related to reference creation, assignment, pinning constraints, and memory moves into atomic-level instructions, including \texttt{BORROW}, \texttt{BIND}, \texttt{DEREF-PIN}, \texttt{DEREF-MOVE}, and \texttt{FORGET}. Figure~\ref{fig:rpil-and-context} demonstrates through an example the RPIL instruction sequences corresponding to each function call and how the program context updates with each instruction's execution. For example, in line 4, since $v_2\!\leadsto\!v_1\in\mathcal{R}$, meaning $v_2$ references $v_1$, the \texttt{DEREF-PIN($v_2$)} instruction causes the referenced target value $v_1$ to enter \texttt{pinned} state. Table~\ref{tab:rpil-instructions} provides a more detailed explanation of the semantics for each RPIL instruction and its impact on the program context.

\begin{table}
\refstepcounter{table}
\label{tab:rpil-instructions}
\centerline {Table 1. Semantics and State Impact of RPIL Instructions.}
\centering\footnotesize
\begin{threeparttable}
\begin{tabular}{p{1.75cm}p{2.95cm}p{2.90cm}}
\toprule
\bfseries Instruction & \bfseries Description & \bfseries Context Impact \\
\midrule
\texttt{BORROW($r,p$)} & Record that $r$ references $p$ & Adds an $r\!\leadsto\!p$ relationship \\\hline
\texttt{BIND($p,q$)} & Binds $q$ to $p$ & Transfers references associated with $q$ to $p$ \\\midrule
\texttt{DEREF-PIN($r$)} & Pin the referenced target if $r\!\leadsto\!p\in\mathcal{R}$ & Sets $p$ to \texttt{pinned} state$^*$ \\\midrule
\texttt{DEREF-MOVE($r$)} & Memory move the referenced target if $r\!\leadsto\!p\in\mathcal{R}$ & If $p$ is now in \texttt{pinned} state, transition it into \texttt{pinned\_moved} state$^*$ \\\midrule
\texttt{FORGET($p$)} & Suppresses the invocation of $p$'s destructor & Sets $p$ to \texttt{forgotten} state$^*$ \\
\bottomrule
\end{tabular}
\begin{tablenotes}
\item[$*$] For detailed behavior, see Section~\ref{sec:pinchecker}.
\end{tablenotes}
\end{threeparttable}
\end{table}

These four aspects of abstraction, by removing nonessential language features and focusing on core semantics directly related to the pin API contract, together form a complete intermediate representation model. that enables accurate description and analysis of pin-related behaviors in Rust programs, providing a clear, operational formal foundation for subsequent static analysis and program synthesis.

\subsection{Extracting Function Behaviors: MIR2RPIL}
To accurately capture the low-level operations in Rust that are related to Pin, we chose MIR~\cite{MIR2025} as the basis for representing function behaviors. MIR is the intermediate representation generated during Rust compilation process. On the one hand, as a lower-level IR, MIR clearly reflects memory operations in a program, including critical behaviors like memory moves. On the other hand, MIR preserves Rust's higher-level semantics, making Pin-related operations identifiable. Due to these two characteristics, MIR not only serves as the foundation for ownership and borrowing checks in the Rust compiler itself, but is also employed in analysis tools like Miri~\cite{Miri2025}. However, MIR representation contains many implementation details irrelevant to pin violation analysis. More importantly, it lacks the ability to directly express key properties like pinning states, which limits its application in our pin violation detection work.

To overcome these limitations, we propose \emph{RPIL (Reference Provenance Intermediate Language)}, a specialized intermediate representation language designed for pin violation detection to represent function behaviors. RPIL eliminates nonessential details from MIR while adding capabilities to track reference targets and pinning state transitions, providing a more suitable foundation for subsequent violation analysis. Based on this, we provide a deterministic translation algorithm from MIR to RPIL that precisely extracts and converts the behaviors of library functions.

\begin{algorithm}
\caption{Translation of MIR to RPIL}
\label{alg:mir2rpil-algorithm}
\KwIn{MIR representation of a library function $\mathcal{F}$}
\KwOut{A set of RPIL variants $\mathcal{T}$}
Initialize translation context $\mathit{ctx}$\;

Push $(\langle\mathcal{F},bb_0\rangle, \mathit{ctx})$ onto worklist $\mathcal{Q}$\;

\While{$\mathcal{Q}$ is not empty}{
    Pop $(\langle\mathit{fn},\mathit{bb}\rangle, \mathit{ctx})$ from $\mathcal{Q}$\;
    \ForEach{$\mathit{stmt}\in\mathit{bb}.\mathit{statements}$}{
        Interpret $\mathit{stmt}$ based on its MIR kind, update $\mathit{ctx}$\;
    }
    \Switch{$\mathit{bb}.\mathit{terminator}$}{
    \uCase{Call($\mathit{fn}'$)}{
    Prepare a new call stack frame in $\mathit{ctx}$\;
    Push $(\langle\mathit{fn}',\mathit{bb}_0\rangle,\mathit{ctx})$ onto $\mathcal{Q}$\;
    }
    \Case{Return}{Retract the top call stack frame from $\mathit{ctx}$\;}
    }
    \If{call stack is empty}{
        Finalize $\mathit{ctx}$ to extract an RPIL sequence, add it to $\mathcal{T}$\;
    }
    \If{$\mathit{bb}.\mathit{terminator}$ has successors}{
        \ForEach{successor basic block $\mathit{bb}_\textit{next}$}{
            Push $(\langle\mathit{fn}, \mathit{bb}_\textit{next}\rangle, \mathit{ctx})$ onto $\mathcal{Q}$\;
        }
    }
}
\Return{$\mathcal{T}$}\;
\end{algorithm}

During the translation process, the system first runs the Rust compiler with the target library's \texttt{Cargo.toml} as a parameter. Once the compiler has generated MIR, the process intercepts the execution. It begins by enumerating all public functions, then for each function, starts sequential translation from the first statement of its first basic block. When reaching a basic block's terminator, if encountering a function call, it enters the first basic block of the called function and continues translation from its first statement; after the function returns, it goes back to the the terminator of the original basic block. If multiple jump targets exist at the end of the current basic block, the translation will attempt each as the next block. The translation continues until encountering a \texttt{return} marker. Throughout the translation process, the system maintains mapping relationships between \emph{local labels} (\textit{i.e.}, MIR \emph{locals}~\cite{MIR2025} such as \texttt{\_1}, \texttt{\_2}, \texttt{\_3}) within the function and state change operations on local labels. Algorithm~\ref{alg:mir2rpil-algorithm} presents the overall flow for translating MIR to RPIL.

MIR and RPIL differ in their models. MIR divides function bodies into multiple basic blocks connected by jump statements, with the execution path determined dynamically at runtime. In contrast, RPIL linearizes all basic blocks during translation, flattening one possible control flow path into one large basic block. Different execution paths yield different RPIL representations, we refer to these different outputs as RPIL \emph{variants}. A single function's MIR may yield multiple RPIL variants by enumerating all possible control-flow paths.

To illustrate the translation process of the MIR2RPIL module, let us consider function \texttt{store\_refs} as a typical case for MIR to RPIL translation. Figure~\ref{fig:rust-and-mir} compares the function's Rust code and its MIR representation (simplified for presentation).

\begin{figure}[!tbp]
  \centering
  \begin{minipage}[t]{0.24\textwidth}
    \begin{lstlisting}[language=Rust,style=colouredRust,basicstyle=\linespread{0.8}\scriptsize\ttfamily]
pub fn store_refs<'a, T>(
  ref1: &'a mut T,
  ref2: &'a mut T
) -> RefStore<'a, T> {
  RefStore {
    store1: Some(ref1),
    store2: Some(ref2),
  }
}
    \end{lstlisting}
  \end{minipage}%
  \begin{minipage}[t]{\textwidth}
    \begin{lstlisting}[language=Rust,style=colouredRust,basicstyle=\linespread{0.8}\scriptsize\ttfamily]
store_refs(_1, _2)
bb0: [
  _3 = Option::Some(copy _1),
  _4 = Option::Some(copy _2),
  _0 = RefStore::<'_, T> {
    store1: move _3,
    store2: move _4
  } -> return
]
    \end{lstlisting}
  \end{minipage}
  \begin{minipage}[t]{\textwidth}
    \begin{lstlisting}[language=Rust,style=colouredRust,basicstyle=\linespread{0.8}\scriptsize\ttfamily]
_0: RefStore<T> = store_refs(_1: &mut T, _2: &mut T)
  BIND(_0[1][1], _1);
  BIND(_0[2][1], _2);
    \end{lstlisting}
  \end{minipage}
  \caption{Rust code (left), MIR (right) and RPIL (bottom).}
  \label{fig:rust-and-mir}
\end{figure}

This function takes two mutable reference parameters and constructs a \texttt{RefStore} structure that encapsulates these two references. From the function's MIR, we can see it contains three basic operations: First, it wraps the first parameter as an \texttt{Option::Some} value and store it in temporary variable \texttt{\_3}; Second, it wraps the second parameter as an \texttt{Option::Some} value and store it in \texttt{\_4}; Third, it combines \texttt{\_3} and \texttt{\_4} into a \texttt{RefStore} instance and assign it to the return value (\texttt{\_0}).

When handling assignments whose right-hand side is a composite data structure, such as in the third step above, MIR2RPIL breaks down the assignment statement into multiple atomic assignments, processing one assignment at a time. For example, when translating \texttt{\_0 = Refstore \{ store1: \_3, store2: \_4 \}}, it splits it into two assignments: \texttt{\_0[1] = \_3} and \texttt{\_0[2] = \_4}. The module then deduces via the mapping relationships, and generates two RPIL instructions: \texttt{BIND(\_0[1][1], \_1); BIND(\_0[2][1], \_2);}.

This algorithm and example highlight three core strategies in the MIR2RPIL module: 1) Control flow linearization: using depth-first traversal of basic blocks to enumerate possible control flow paths, converting branch structures into sequential execution paths; 2) Structured decomposition during assignment: breaking down assignments with composite right-hand sides into atomic assignment operations, ensuring each RPIL instruction operates on only a single label; 3) Tracking of state changes: capturing the functions' modifications to values' states, accurately recording key state-changing operations like \texttt{DEREF-PIN} and \texttt{DEREF-MOVE}. These strategies effectively extract critical function behaviors from MIR, providing an accurate behavior model for subsequent constraint solving in detecting pin violations.

\subsection{Synthesizing Pin Violations: PinChecker}\label{sec:pinchecker}
The PinChecker detection tool is the core component in our work for detecting pin API contract violations. Its fundamental idea is to construct a program by composing multiple library functions, such that the program reaches a pin violation state after execution (\textit{e.g.}, when a pinned value undergoes a memory move). To achieve this goal, PinChecker implements a special reversible interpreter, that supports both forward execution of programs to determine their final state and reverse execution that that performs backward reasoning from a desired state to synthesize a program capable of reaching that state.

\subsubsection{Overview}
PinChecker synthesizes pin-violating programs by establishing a many-to-one mapping between programs and their \emph{post-execution contexts}. It takes as input the signatures and RPIL instructions of library functions, then employs a Prolog-based reversible interpreter to perform bidirectional reasoning. PinChecker leverages this property by querying the interpreter to derive programs that result in contexts exhibiting pin violations. When such a program is successfully synthesized, it confirms that the provided set of library functions can be composed to violate the pin contract.

PinChecker is implemented as a suite of utility predicates that together capture the correlation between a program and its execution context. Each predicate is designed to address a specific aspect of the analysis; for example, \texttt{borrow/3} encodes the reference relationships between values, and \texttt{status/3} tracks values' pin/forget states.

Figure~\ref{fig:forward-and-backward} presents examples of PinChecker's predicates in use. The top example demonstrates forward interpretation, while the latter two illustrate backward interpretation for program synthesis. Because multiple programs can satisfy the same constraints, backward interpretation may yield multiple solutions, as shown in the middle example.

\begin{figure}[!tbp]
  \centering
  \begin{lstlisting}[language=Rust,style=colouredRust,basicstyle=\linespread{0.8}\footnotesize\ttfamily]
?- Prgm = [
  v1 = SelfRef::new(),     // ;
  v2 = borrow_mut(v1),     // BORROW(v2, v1);
  v3 = mylib::pin_new(v2), // DEREF-PIN(v2);
  v4 = deref_move(v2),     // DEREF-MOVE(v2);
], <@\textcolor{ForestGreen}{borrows}@>(Prgm, v2, <@\textbf{X}@>), <@\textcolor{ForestGreen}{state}@>(Prgm, v1, <@\textbf{Y}@>).
<@\textbf{X}@> = v1,
<@\textbf{Y}@> = pinned_moved.

?- length(Prgm, 2), <@\textcolor{ForestGreen}{borrows}@>(<@\textbf{Prgm}@>, v2, v1).
<@\textbf{Prgm}@> = [
  v1 = SelfRef::new(),
  v2 = borrow(v1),
];
<@\textbf{Prgm}@> = [
  v1 = SelfRef::new(),
  v2 = borrow_mut(v1),
].

?- length(Prgm, 4), <@\textcolor{ForestGreen}{state}@>(<@\textbf{Prgm}@>, <@\textbf{X}@>, pinned_moved).
<@\textbf{Prgm}@> = [
  v1 = SelfRef::new(),
  v2 = borrow_mut(v1),
  v3 = mylib::pin_new(v2),
  v4 = deref_move(v2),
],
<@\textbf{X}@> = v1.
  \end{lstlisting}
  \caption{Usage examples of PinChecker's Prolog predicates (marked green). The top example demonstrates forward interpretation, while the bottom two illustrate backward interpretation for program synthesis.}
  \label{fig:forward-and-backward}
\end{figure}

Table~\ref{tab:core-predicates} lists the core predicates used in PinChecker. The predicates work cohesively to guide the synthesis process, pruning the search space of candidate programs.

\begin{table*}[!tbp]
\refstepcounter{table}
\label{tab:core-predicates}
\centerline {Table 2. Core predicates in PinChecker's implementation.}
\centering\footnotesize
\begin{tabular}{p{6.1cm}p{7.6cm}cc}
\hline
\bfseries Signature & \bfseries Description & \bfseries Directionality & \bfseries Nature  \\
\hline
\texttt{typeof(Prgm,V,T)} & Variable \texttt{V} in program \texttt{Prgm} is of type \texttt{T} & Bidirectional & Synthetic \\\hline
\texttt{liveness(Prgm,V,L)} & Variable \texttt{V} in program \texttt{Prgm} is \texttt{L} (\texttt{alive}/\texttt{dead}) & Bidirectional & Synthetic \\\hline
\texttt{borrows(Prgm,R,P)} & Reference relationship $\texttt{R}\leadsto\texttt{P}$ exists in program \texttt{Prgm} & Bidirectional & Synthetic \\\hline
\texttt{state(Prgm,P,St)} & Value \texttt{P} in program \texttt{Prgm} is in state \texttt{St} & Bidirectional & Synthetic \\\hline
\texttt{fn\_type(Fn,[Ty1,Ty2,...],RetTy)} & Function \texttt{Fn}'s type signature is $\texttt{Ty1}\times\texttt{Ty2}\times\cdots\to\texttt{RetTy}$ & Bidirectional & Inherited \\\hline
\texttt{fn\_rpil(Fn,[Ret,Arg1,Arg2,...],Rpil)} & Gets the RPIL of function \texttt{Fn}, replacing placeholders for return value (\texttt{\_0}) and arguments (\texttt{\_1}, \texttt{\_2}, \ldots) with concrete variables & Forward-only & Inherited \\
\hline
\end{tabular}
\end{table*}

\subsubsection{Constraint Rules Guiding PinChecker}\label{sec:rules}
PinChecker implements a set of formal rules, capturing key aspects of the type system, variable liveness, reference relationships, and state transitions. These rules ensure correct interaction between predicates during program synthesis.

\begin{pcrule}[Variable Well-typedness]\label{rule:well-typedness}
Let the last line in program be \texttt{Ret = Fn(Arg1,$\ldots$,Argk)}. The variable \texttt{Ret} is well-typed (\textit{i.e.}, clause \texttt{typeof(Prgm, Ret, \_)} is satisfiable), if all of the following conditions are satisfied.
\begin{enumerate}[label=\arabic*.]
\item Variables \texttt{Arg1}, \ldots, \texttt{Argk} are well-typed and are alive;
\item \texttt{Fn}'s type signature $\texttt{Ty1}\times\cdots\times\texttt{Tyk}\to\texttt{RetTy}$ is satisfiable;
\end{enumerate}
\end{pcrule}

Rule~\ref{rule:well-typedness} establishes the typing requirements for function calls. It ensures that all arguments are properly typed and available at the point of function invocation.

\begin{pcrule}[Variable Liveness]\label{rule:liveness}
Let the last line in program be \texttt{Ret = Fn(Arg1,$\ldots$,Argk)}. The variable \texttt{V}'s liveness will be changed by the last line, if any of the following conditions is satisfied.
\begin{enumerate}[label=\arabic*.]
\item Variable \texttt{V} turns \texttt{alive}, when \texttt{V = Ret} and \texttt{Ret} is well-typed;
\item Variable \texttt{V} turns \texttt{dead}, when \texttt{V} is one of \texttt{Arg1}, \ldots, \texttt{Argk}, and function \texttt{Fn} can consume \texttt{V};
\end{enumerate}
\end{pcrule}

Rule~\ref{rule:liveness} manages variable lifetimes by tracking value consumption. A variable becomes alive when assigned a well-typed return value, and becomes dead after being used in function arguments. The only two exceptions are \texttt{borrow} and \texttt{borrow\_mut}, which create references without consuming their arguments.

\begin{pcrule}[Reference Introduction]\label{rule:reference-introduction}
Let \texttt{Rpil} denote the RPIL sequence of the function that the last program line calls. The reference relationship $\texttt{R}\!\leadsto\!\texttt{P}$ is introduced, if any of the following conditions is satisfied.
\begin{enumerate}[label=\arabic*.]
\item \texttt{BORROW(Lhs, Rhs)} is one instruction in \texttt{Rpil}, where \texttt{Lhs} evaluates to \texttt{R}, and \texttt{Rhs} evaluates to \texttt{P};
\item \texttt{BIND(Lhs, Rhs)} is one instruction in \texttt{Rpil}, where reference relationship $\textrm{Rep}_{\texttt{Lhs}\to\texttt{Rhs}}(\texttt{R})\leadsto\texttt{P}$ holds;
\end{enumerate}
$\textrm{Rep}_{v'\to v}(\cdot)$ is an operator to replace a value's belonging variable $v'$ with $v$.
\end{pcrule}

\begin{pcrule}[Reference Retraction]\label{rule:reference-retraction}
The reference relationship $\texttt{R}\leadsto\texttt{P}$ does not hold any more, when the variable that \texttt{R} belongs to turns \texttt{dead}, or when the variable that \texttt{P} belongs to turns \texttt{dead}.
\end{pcrule}

Rules~\ref{rule:reference-introduction} and \ref{rule:reference-retraction} form the core mechanism for tracking reference relationships. These rules implement the semantics of RPIL instructions \texttt{BORROW} and \texttt{BIND}. The \texttt{BORROW} instruction introduces a new reference relationship, while \texttt{BIND} transfers existing references between values. For example, if $\texttt{v2[1]}\leadsto\texttt{v1}$ holds, then executing \texttt{BIND(v3, v2)} introduces $\texttt{v3[1]}\leadsto\texttt{v1}$. The \texttt{BORROW} operation supports dereferencing through existing relationships: \texttt{BORROW(v4, (*v2[1]))} would first resolve $\texttt{*v2[1]}$ to \texttt{v1} via the existing reference, then introduce $\texttt{v4}\leadsto\texttt{v1}$.

\begin{figure}[!tbp]
    \centering
    \includegraphics[width=\linewidth]{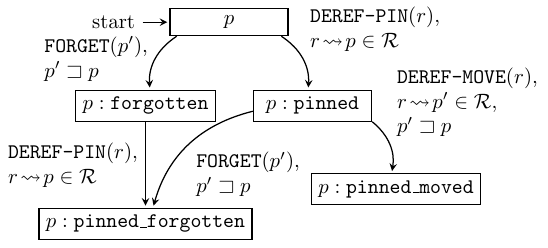}
    \caption{State transition of a value.}
    \label{fig:state-transition}
\end{figure}

\begin{pcrule}[State Transition of Values]\label{rule:state-transition}
A value can be in initial state, or one of \texttt{pinned}, \texttt{forgotten}, \texttt{pinned\_moved} and \texttt{pinned\_forgotten}. The state transitions for a value are shown in Figure~\ref{fig:state-transition}, with transition conditions labeled on the arrows.
\end{pcrule}

Take the transition from \texttt{pinned} to \texttt{pinned\_moved} as an example, the transition requires a \texttt{DEREF-MOVE($r$)} instruction, along with the conditions that $r\leadsto p'\in\mathcal{R}$ and $p'\sqsupset p$ hold, meaning a reference relation from $r$ to $p'$ exists in the program context's reference relationship graph, and $p'$ is a sub-location of $p$. For instance, if \texttt{$v_1$[1]} is in \texttt{pinned} state and $v_2$ holds a reference to $v_1$, then executing \texttt{DEREF-MOVE($v_2$)} will cause \texttt{$v_1$[1]}'s state to transition to \texttt{pinned\_moved}. This corresponds to using \texttt{mem::swap} through a reference to a value causes that referenced value and its internal sub-places to undergo a memory move.

Rule~\ref{rule:state-transition} forms the core mechanism for tracking values' states pertaining Rust's pinning APIs, especially pin violations. The latter two states, \texttt{pinned\_moved} and \texttt{pinned\_forgotten} especially represent pin violating states.

\subsubsection{Synthesizing Pin Violations}\label{sec:pinchecker-violations}
The two categories of pin violations correspond to \texttt{pinned\_moved} and \texttt{pinned\_forgotten}. When any value in the program's execution context enters either of these states, it indicates that the program has exhibited behavior that violates the pin API contract.

PinChecker's program synthesis mechanism is based on constraint solving: it searches for function call sequences that satisfy specific constraint conditions by running the interpreter in reverse. Specifically, the target state is defined as a program context exhibiting a pin violation, described with a clause like \texttt{state(Prgm, \_, pinned\_moved)} (where the wildcard ``\texttt{\_}'' is a Prolog feature denoting existential constraint), requiring at least one value to be in this anomalous state after program \texttt{Prgm}'s execution. Given the target context, the interpreter uses a constraint-based backward interpretation strategy to synthesize program \texttt{Prgm} line by line from back to front, searching for all possible programs that can reach this state. If at least one program is successfully generated, it demonstrates that the combination of related functions in the target library can give rise to a pin violation, proving the existence of safe abstraction defects in the library.

PinChecker is guaranteed to terminate. If the constraint solving space is exhausted without finding a feasible solution, the system will terminate and return an empty result set, ensuring the analysis will not run indefinitely.

\section{Implementation and Evaluation}
We have implemented PinChecker, a tool that automatically detects pin contract violations caused by improper safe abstractions in Rust libraries through constraint solving based program synthesis. To systematically evaluate PinChecker, we conducted experiments on real Rust crates from crates.io, focusing on the following three research questions:

\begin{enumerate}[label=\textbf{RQ\arabic*.},left=0pt]
\item Can PinChecker detect pin contract violations in real-world Rust libraries?
\item How does PinChecker perform in synthesizing pin violation programs for libraries of various scales and complexities?
\item How expressive is PinChecker's linear program form in capturing real-world pinning API usage?
\end{enumerate}

\subsection{Implementation of MIR2RPIL}
MIR2RPIL extends the Rust compiler through the \texttt{rustc\_private} interface to extract and convert MIR representations of all public (pub) functions into RPIL. The implementation operates in two stages:

In the first stage, it obtains the compilation context by invoking Cargo commands \cite{Cargo2025} with the target crate's Cargo.toml, then identifies the arguments to the last \texttt{rustc} command, the command that compiles the target crate itself.

In the next stage, MIR2RPIL replaces \texttt{rustc} to execute the last command, using the same command-line arguments identified in the previous stage. MIR2RPIL injects its logic during the \emph{AfterAnalysis} phase, traversing the crate to filter public function definitions. For each function, it applies Algorithm~\ref{alg:mir2rpil-algorithm} to translate their MIR into RPIL.

\subsection{Implementation of PinChecker}\label{sec:impl-pinchecker}
\begin{figure}[!tbp]
  \centering
  \includegraphics[width=\linewidth]{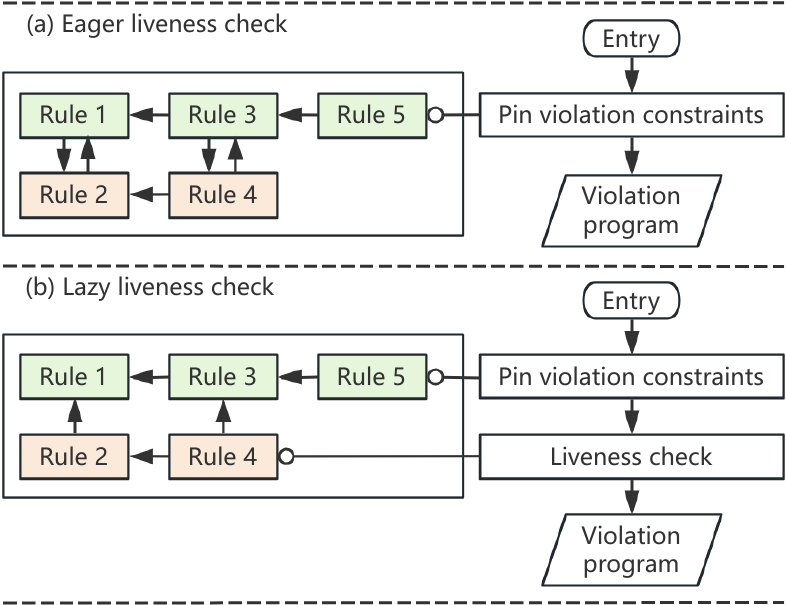}
  \caption{Comparison between PinChecker's implementation strategies.}
  \label{fig:pinchecker-variants}
\end{figure}

PinChecker is implemented in Prolog as a reversible interpreter-based violation detection tool using the core predicates listed in Table~\ref{tab:core-predicates} and auxiliary predicates that enforce the five constraint rules in Section~\ref{sec:rules}. We use SICStus Prolog~\cite{SICStusProlog2024} for execution. Our initial ``monolith'' implementation, integrated all five rules, but we found that liveness rules (Rule~\ref{rule:liveness} and Rule~\ref{rule:reference-retraction}) were time-consuming to evaluate. We therefore developed a variant with \emph{lazy liveness check} that defers liveness verification. Figure~\ref{fig:pinchecker-variants} compares the two implementation strategies. We evaluate their performance in Section~\ref{sec:rq2}.

The synthesis procedure in PinChecker requires pre-determining the target program length. Accordingly, PinChecker begins by synthesizing 1-line programs. If no violation program is generated at the current length, the synthesis length limit is iteratively increased by one. Upon finding the first violation program, PinChecker reports the result and terminates further exploration.

\subsection{Selection of Libraries and API Functions}
To evaluate PinChecker's effectiveness, we selected a representative set of Rust libraries from crates.io for testing. We primarily focused on libraries related to the Pin mechanism, as these libraries are widely used in asynchronous programming and self-referential data structures and have significant influence on the Rust ecosystem. We initially performed a screening on crates.io with keywords including ``pin'', ``unpin'', and ``self-reference''. During the selection process, we followed these criteria:

\begin{enumerate}[label=\arabic*.,left=0pt]
\item \textbf{Pure Rust Implementation.} The selected libraries must be written in Rust, as our tool needs to extract MIR and transform it into RPIL for analysis.
\item \textbf{Pin-related Safe Abstraction Functions.} Libraries contain public functions that provide safe abstractions over unsafe pinning APIs.
\item \textbf{Pinnable Data Structures Implementing the Drop Trait.} Libraries contain pinnable (types which are not \texttt{Unpin}) data type definitions that implement the \texttt{Drop} trait.
\end{enumerate}

Following these criteria, we ultimately selected 13 qualifying libraries. For each library, we first run MIR2RPIL to list the paths, names, and type signatures of all public (pub) functions along with their RPIL translation results. PinChecker allows for the selection of up to 10 functions to be analyzed. To ensure comprehensive coverage, the public functions listed by the MIR2RPIL module are sorted, prioritizing safe abstraction functions containing internal \texttt{unsafe} blocks. The selection strategy is: prioritize safe abstraction functions (\textit{i.e.}, the first 10 functions) for testing, while allowing users to adjust selections based on testing requirements, such as when users want to test specific modules. In addition to library functions, we added 4 default functions to each library's function set, including 2 for supporting Rust borrowing operations: \texttt{\&} and \texttt{\&mut}, 1 for memory move operations: \texttt{deref\_move} (such as \texttt{mem::swap} or \texttt{mem::replace}), and 1 for the ``drop-forgetting'' operation: \texttt{forget}.

\subsection{Experiment Setup}
PinChecker attempts to automatically synthesize potential pin violation triggering cases within a time limit. For efficiency and comprehensiveness, we ran experiments in parallel in a server environment to reduce both individual library detection time and overall detection time. During the experiments, we used the following hardware and software environment. Hardware environment: We used a server equipped with an Intel Xeon Gold 6326 processor and 256GB of memory. Software environment: Ubuntu 23.10 operating system. The Rust compiler version was 1.87.0-nightly to ensure compatibility with the latest language features and library versions. We used SICStus Prolog 4.9.0 to run PinChecker. During the experiments, the server ran all analysis tasks concurrently in containers.

For each crate, we first ran the MIR2RPIL module to extract the RPIL representations of the selected functions from the library; these representations were then incorporated into a PinChecker code template to form a complete Prolog program. Next, we ran PinChecker to analyze the combinations of these functions. To control experiment scale and duration, we set a runtime limit of 10 hours per crate.

It is worth noting that while experiments were conducted on a high-performance computer, this was primarily to enable parallel search and reduce runtime. For analyzing individual libraries, the computational requirements are relatively low; an ordinary personal computer with lower computational resources suffices.

\section{Evaluation Results}
In this section, we analyze PinChecker's detection results and performance to address RQ1--RQ3.

\subsection{RQ1: Effectiveness in Detecting Pin Violations}
To address RQ1, we evaluated PinChecker's effectiveness in detecting pinning API contract violations in real-world Rust libraries. We selected a diverse set of libraries that are widely used and involve pinning APIs, as shown in Table~\ref{tab:dataset-overview}. The dataset is sorted from the highest number of public function definitions to the lowest. All these libraries have significant numbers of downloads and exhibit high percentages of unsafe code. Even those with comparatively lower unsafe code percentages (less than 5\%) primarily use unsafe pinning API calls within.

\begin{table*}[!tbp]
  \refstepcounter{table}
  \label{tab:dataset-overview}
  \centerline {Table 3. Characteristics and pinning API usage patterns in evaluated Rust libraries.}
  \centering\small
  \begin{tabular}{|l|c|c|c|r|r|}
    \hline
    \bfseries Library & \bfseries \CellWithForcedBreak{\# Pub Fn Definitions\\/ \# Fn Definitions} & \bfseries \CellWithForcedBreak{\# Unsafe Pin API calls\\+ \# Drop Trait Impls} & \bfseries \CellWithForcedBreak{\# Unsafe Lines\\/ \# Total Lines} & \bfseries \CellWithForcedBreak{Unsafe\\Percentage} & \bfseries All-time Downloads \\
    \hline
moveit          & 38 / 115 & 26 + 6 &  407 / 1,211 & 33.61\% &   741,454 \\\hline
pin-init        & 36 /  98 & 17 + 5 &  101 /   880 & 11.48\% &    23,625 \\\hline
rio             & 31 /  95 &  0 + 7 &  187 / 2,143 &  8.71\% & 3,777,002 \\\hline
pin-rc          & 29 /  55 &  3 + 0 &   24 /   364 &  6.59\% &     1,421 \\\hline
owned-pin       & 21 / 121 &  5 + 2 &  154 / 1,030 & 14.95\% &     9,406 \\\hline
pin-queue       & 20 /  49 &  8 + 0 &  101 /   462 & 21.86\% &     1,693 \\\hline
pin-cell        & 15 /  23 &  2 + 0 &    2 /   112 &  1.40\% &    59,544 \\\hline
recycle-box     & 11 /  46 &  4 + 2 &   81 /   491 & 16.50\% &    28,414 \\\hline
pin\_array      &  9 /  22 &  7 + 0 &   11 /   199 &  5.53\% &     2,822 \\\hline
pinned\_vec     &  8 /  24 &  2 + 0 &    2 /   202 &  0.99\% &    11,797 \\\hline
lazy-pinned     &  7 /   8 & 18 + 0 &   17 /    84 & 20.24\% &     1,444 \\\hline
pinus           &  4 / 103 & 26 + 2 &  564 / 1,185 & 47.59\% &    12,993 \\\hline
assert-unmoved  &  2 /  71 & 11 + 1 &   12 /   619 &  1.72\% &     7,100 \\
    \hline
  \end{tabular}
\end{table*}

We then executed PinChecker on each crate, with a maximum runtime limit of 10 hours per crate. Table~\ref{tab:pinchecker-results} summarizes the detection results. The columns denote the library name, the maximum target synthesis length attempted, and the number of stub programs explored during synthesis. Libraries identified with violations are marked with a $\star$ symbol.

\begin{table}[!tbp]
  \refstepcounter{table}
  \label{tab:pinchecker-results}
  \centerline {Table 4. PinChecker synthesis results for each crate.}
  \centering\small
  \begin{threeparttable}
  \begin{tabular}{|l|c|@{\hskip 2.2em}c|}
    \hline
    \bfseries Library & \bfseries Max. Length & \bfseries \# Synthesized\phantom{$^*$} \\
    \hline
moveit\,$\star$1 &  8 &        839,579$^*$ \\\hline
pin-init         &  9 &  1,430,742,432\phantom{$^*$} \\\hline
rio\,$\star$2    &  7 &        204,749$^*$ \\\hline
pin-rc           &  9 &  1,033,101,237\phantom{$^*$} \\\hline
owned-pin        &  9 &    974,834,349\phantom{$^*$} \\\hline
pin-queue        & 10 & 12,710,367,475\phantom{$^*$} \\\hline
pin-cell         &  9 &  1,146,865,944\phantom{$^*$} \\\hline
recycle-box      &  9 & 12,041,352,216\phantom{$^*$} \\\hline
pin\_array       &  9 &  1,082,026,229\phantom{$^*$} \\\hline
pinned\_vec      &  9 &  1,566,995,155\phantom{$^*$} \\\hline
lazy-pinned      &  9 & 10,666,622,602\phantom{$^*$} \\\hline
assert-unmoved   &  9 &  2,330,453,733\phantom{$^*$} \\
    \hline
  \end{tabular}
  \begin{tablenotes}
  \item[$*$] Exploration terminates upon finding the first bug-revealing program.
  \end{tablenotes}
  \end{threeparttable}
\end{table}

Table~\ref{tab:bugs-caught} presents a closer look at the bugs detected. We will discuss these bugs in detail below.

\begin{table}[!tbp]
  \refstepcounter{table}
  \label{tab:bugs-caught}
  \centerline {Table 5. Bugs caught by PinChecker.}
  \centering\footnotesize
  \begin{tabular}{|l|l|r|r|l|}
    \hline
    \bfseries $\star$ & \bfseries Bug Type & \bfseries \CellWithForcedBreak{Min. Lines\\to Induce} & \bfseries \CellWithForcedBreak{Time to\\Discovery} & \bfseries Confirmed? \\
    \hline
$\star$1 & Move-After-Pin &  8 &  2.4s & Yes \\\hline
$\star$2 & Pin-And-Leak   &  7 &  0.6s & Yes \\
    \hline
  \end{tabular}
\end{table}

Bug $\star$1 is a first type of pin violation (Move-after-Pin) in the moveit library, where a value that has already been pinned is subsequently moved. The minimal violation program generated by the tool is shown in Figure~\ref{fig:violation-moveit}. Through a combination of eight function calls, the program gradually bypasses Pin's protection: First, line 2 creates Pin pointer \texttt{v2}, which pins its referent (\texttt{v1}), putting it in \texttt{pinned} state. Then, lines 3--6 construct intermediate variable v5, preparing parameters needed for line 6's function call. Next, lines 6--7 transfer the reference relationship to \texttt{v1} first into \texttt{v6}'s interior, then extract it as \texttt{v7}. Finally, line 8 performs a memory move on \texttt{v7}'s referent (\texttt{v1}) through \texttt{mem::swap}, causing it to enter \texttt{pinned\_moved} state, triggering a pin violation.

\begin{figure}[!tbp]
  \centering
  \begin{lstlisting}[language=Rust,style=colouredRust,basicstyle=\scriptsize\ttfamily]
let mut v1 = Unmovable::new();
let mut v2 = Box::pin(v1);
let mut v3 = SlotDropper::new();
let mut v4 = &mut v3;
let mut v5 = SlotDropper::new_unchecked_hygine_hack(v4);
let mut v6 = <_ as moveit::DerefMove>::deref_move(v2, v5);
let mut v7 = &mut *v6;
let mut v8 = mem::swap(v7, &mut Unmovable::new());
----------------------------------------------------------
line 1: ;
line 2: BIND(v2[1], v1); BORROW(v2, v2[1]); DEREF-PIN(v2);
line 3: ;
line 4: BORROW(v4, v3);
line 5: BORROW(v5[1], (*v4)[1]);
line 6: BORROW(v6[1], (*v2));
line 7: BIND(v7, v6[1]);
line 8: DEREF-MOVE(v7);
  \end{lstlisting}
  \caption{Minimal violation program for moveit v0.5.1 generated by PinChecker, with RPIL instructions for each line.}
  \label{fig:violation-moveit}
\end{figure}

The vulnerability originates from the library function \texttt{new\_unchecked\_hygine\_hack}, which exposes a means of obtaining a mutable reference. After confirming the issue, developers removed this function in v0.6.0.

Bug $\star$2 is a second type of pin violation (Pin-And-Leak) in the rio library, violating the Drop guarantee. The minimal violation program is shown in Figure~\ref{fig:violation-rio}. This issue arises from creating Pin pointers through \texttt{rio::Completion} struct's safe abstraction interface, then calling the standard library's \texttt{mem::forget} to suppress drop execution. This vulnerability can lead to use-after-free memory safety issues by accessing freed buffers while kernel I/O operations are incomplete. This issue has been recorded in vulnerability exposure databases as RUSTSEC-2020-0021~\cite{CVERio2020}.

\begin{figure}[!tbp]
  \centering
  \begin{lstlisting}[language=Rust,style=colouredRust,basicstyle=\scriptsize\ttfamily]
let mut v1 = rio::new();
let mut v2 = io::Result::unwrap(v1);
let mut v3 = &*v2;
let mut v4 = io::stdin();
let mut v5 = &v4;
let mut v6 = Default::default();
let mut v7 = &v6;
let mut v8 = Default::default();
let mut v9 = rio::Uring::read_at(v3, v5, v7, v8);
let mut v10 = &mut v9;
let mut v11 = pin::Pin::new(v10);
let mut v12 = mem::forget(v9);
  \end{lstlisting}
  \caption{Minimal violation program for rio v0.9.4.}
  \label{fig:violation-rio}
\end{figure}

\subsection{RQ2: Performance in Pin Violation Synthesis}\label{sec:rq2}
\begin{figure}[!tbp]
  \centering
  \includegraphics[width=\linewidth]{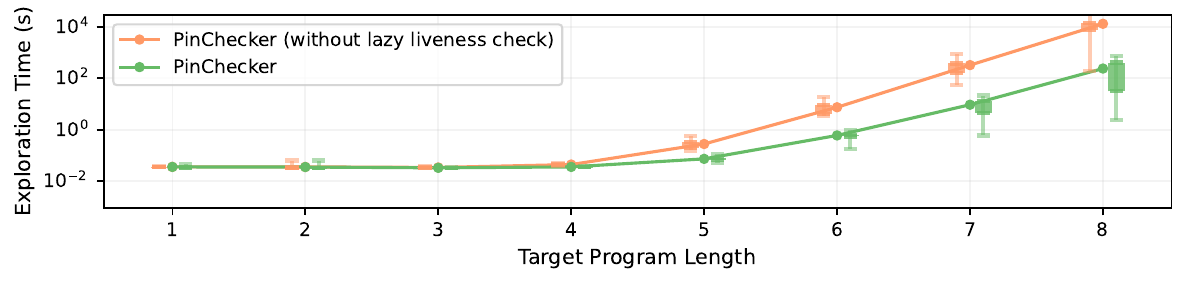}
  \caption{Comparison between PinChecker variants' average exploration time. The candlestick beside each data point shows the 25th and 75th percentiles.}
  \label{fig:benchmark-comparison}
\end{figure}

To address RQ2, we further examine PinChecker's performance, comparing both PinChecker variants (Section~\ref{sec:impl-pinchecker}), running each on all libraries. For each library, we instructed them to search for violation programs with target synthesis lengths ranging from 1 up to 8, and we recorded the time taken for exploration until termination. Figure~\ref{fig:benchmark-comparison} presents the experimental results, illustrating the mean exploration times along with the 25th and 75th percentile bounds for each target program length. The results reveal that the lazy liveness check variant of PinChecker consistently outperforms the monolith variant, and the performance advantage increases as the target program length increases.

\begin{figure}[!tbp]
  \centering
  \includegraphics[width=\linewidth]{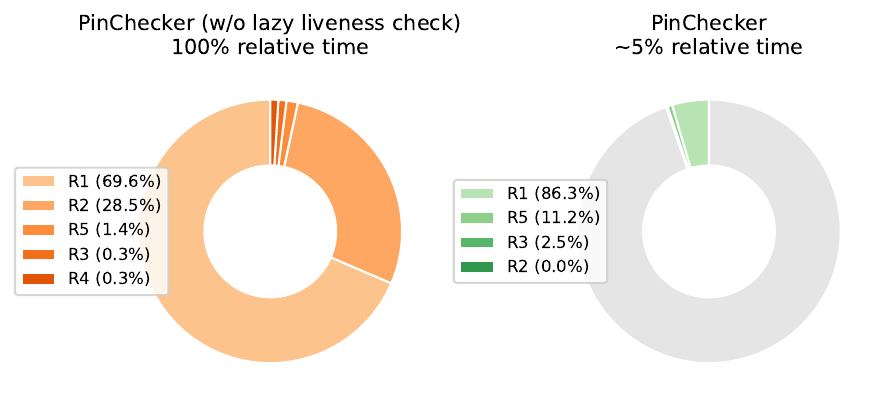}
  \caption{Comparison between PinChecker variants' rule composition.}
  \label{fig:rule-composition}
\end{figure}

We further instrumented the program to analyze how much time is spent on each of the five constraint rules. Specifically, we ran both PinChecker variants on each library with target length set to 5, collected instrumentation data, and calculated the average time spent on each rule. The results are shown in Figure~\ref{fig:rule-composition}. With lazy liveness checking, PinChecker is able to perform $\sim$19.1 times as fast as the version without lazy liveness checking, needing only $\sim$5\% of runtime. Moreover, given the same amount of execution time, with lazy liveness checking, PinChecker is able to allocate higher proportion of computation (11.2\% compared to 1.4\%) to evaluating Rule~\ref{rule:state-transition}, which is the core rule that PinChecker relies on to detect pin violations.

\subsection{RQ3: Expressiveness of Linear Programs}
To address RQ3, we evaluate the effectiveness of PinChecker's linear program synthesis approach by examining how well it can represent real-world Pin API usage patterns. We studied usage examples for the crates in our evaluation from two sources: 1) dependent crates using the target crates (18 examples), and 2) the target crates' official documentation and test cases (12 examples). Table~\ref{tab:linear-form} shows the distribution across libraries. Under this criterion, we collected a total of 30 distinct usage examples. For each example, our analysis procedure involved three steps: 1) we identify invocations of the crate's pin-related APIs, forming a self-contained API call sequence; 2) we rewrite each API call sequence in linear form while maintaining their original semantics; 3) we verify semantic equivalence through MIR comparison and manual inspection.

\begin{table}[!tbp]
  \refstepcounter{table}
  \label{tab:linear-form}
  \centerline {Table 6. Expressiveness evaluation dataset composition.}
  \centering\footnotesize
  \begin{tabular}{|l|c|c|r|}
    \hline
    \bfseries Crate Category & \bfseries \# Examples & \bfseries \# Linearizable & \bfseries \% \\
    \hline
    Self-referential struct & 15 & 14 & 93\%\\\hline
    Async primitives        &  9 &  8 & 89\%\\\hline
    Pin-safe data structure &  6 &  5 & 83\%\\\hline
    \textbf{Total}          & \textbf{30} & \textbf{27} & \textbf{90\%} \\
    \hline
  \end{tabular}
\end{table}

Our analysis revealed that 90\% (27 out of 30) of the real-world Pin-related API use cases could be successfully expressed in PinChecker's linear form without compromising semantics. The 3 cases that could not be adequately expressed in linear form all involved closure-based APIs. Figure~\ref{fig:closure-example} shows a characteristic failure case involving closure parameters. These APIs require function pointer types as input parameters, which would necessitate extending our model to support multiple function definitions rather than the current single-line program flow approach.

\begin{figure}[!tbp]
\begin{lstlisting}[language=Rust,style=colouredRust,basicstyle=\scriptsize\ttfamily]
let ret = moveit::new::by_raw(unsafe |space| {
  let space = Pin::into_inner_unchecked(space).as_mut_ptr();
});
\end{lstlisting}
\caption{An example involving a closure parameter.}
\label{fig:closure-example}
\end{figure}

These findings suggest that PinChecker's approach to program synthesis, while constrained to linear form, remains sufficiently expressive to model and analyze the vast majority of real-world Pin API usage patterns. This expressiveness validates our design decision to employ a simplified program representation that facilitates efficient constraint solving while maintaining practical applicability to real-world code.

\subsection{Limitations}
While PinChecker effectively detects pin API contract violations, some limitations remain that suggest areas for future enhancement.

\subsubsection{Extending Safety Feature Coverage}
The current implementation of PinChecker focuses primarily on the Pin mechanism. A natural extension would be to broaden the analysis scope to cover more Rust safety issues related to \texttt{unsafe} code usage. We believe that extending RPIL's instruction set and enhancing the interpreter's semantic rules would enable more safety analyses.

\subsubsection{Supporting Macro-based Abstractions}
While PinChecker successfully analyzes function-level pinning API safety abstractions, some Rust libraries implement pinning API abstractions through procedural macros. The code generation nature of macro expansion introduces additional complexity to static analysis. We believe that incorporating post-macro-expansion code analysis will offer a more comprehensive coverage.

\subsubsection{Optimizing Analysis Performance}
While most crates achieve satisfactory results within acceptable time for analysis, complex API functions with numerous parameters and libraries with intricate interactions can be time-consuming. Future work could explore more sophisticated constraint solving techniques, such as state pruning mechanisms, to further enhance analysis efficiency.

\section{Related Work}
This research intersects multiple domains including safe encapsulation of unsafe Rust code and constraint solving based program synthesis. In this section, we review the related work and discuss their connection to our research.

\textbf{Sound Usage of Unsafe Rust Code.} Ensuring the soundness of safe abstractions over unsafe code is crucial for maintaining Rust's safety guarantees. Rao~\cite{VerifyInteriorUnsafe2024} proposed unsafety isolation graphs to model the usage and encapsulation of unsafe code, facilitating soundness checking of unsafe code encapsulation. Other research has investigated how to safely use existing unsafe code. Liu \textit{et al.}~\cite{XRust2020} presented XRust, which mitigates security threats from unsafe code by ensuring the integrity of data flow from unsafe Rust code to safe Rust code. Almohri and Evans~\cite{IsolateUnsafe2018} introduced Fidelius Charm to protect designated memory data from unauthorized access through unsafe libraries. Lamowski \textit{et al.}~\cite{SandboxUnsafe2017} presents a sandboxing solution that isolates unsafe C library code in separate processes, protecting the main Rust program from memory corruption in unsafe code.

\textbf{Constraint Solving Based Program Synthesis.} Fiala \textit{et al.}~\cite{LevRustTypes2023} introduced RusSOL, a tool for synthesizing programs that satisfy functional specifications while conforming to Rust's type system. Their approach leverages Synthetic Ownership Logic to derive programs that comply with Rust's type system, benefiting from the type system to reduce search space. Similarly, Takashima \textit{et al.}~\cite{SyRust2021} presented SyRust, a framework for automatically synthesizing well-typed Rust programs to test unsafe library APIs. SyRust draws inspiration from H+~\cite{Hplus2019} and SyPet~\cite{FengPOPL17}, and extends their approaches by encoding Rust's ownership system type constraints into logical forms, guiding the program synthesis.

\textbf{Logic Programming Based Program Synthesis.} Kosarev \textit{et al.}~\cite{RelationalSynthesis2020} explored a declarative approach using relational programming to synthesize implementations of pattern matching constructs. By representing both high-level semantics and intermediate implementation language semantics relationally, they enabled extensible synthesis of new language feature implementations. Byrd \textit{et al.}~\cite{FunctionalPearl2017} demonstrated how to write relational interpreters in logic programming languages, enabling both program execution and synthesis from execution traces. In the context of Rust, Dewey \textit{et al.}~\cite{CLPT2015} applied constraint logic programming (CLP) to generate well-typed programs, with a focus on testing the Rust type checker.

\section{Conclusion}
We have presented PinChecker, an automatic detection tool for unsound safe abstractions of Rust's unsafe pinning APIs. Our approach extracts function behaviors into a novel intermediate representation, RPIL, which accurately models references and pinning state transitions. By employing a reversible interpreter implemented in Prolog, PinChecker is able to both emulate forward program execution and perform backward synthesis to generate minimal counterexample programs that trigger pin contract violations. Our evaluation on real-world Rust libraries demonstrates PinChecker's ability to generate counterexample programs and find 2 confirmed issues. In future work we plan to extend PinChecker's support to a broader subset of Rust's unsafe functionalities. We also plan to improve the scalability of our constraint solving through more advanced constraint solving techniques.

\section*{Acknowledgment}
This work was supported in part by the National Natural Science Foundation of China (NSFC) under Grant No. 62272220.

\bibliographystyle{ieeetr}
\bibliography{base}











\balance

\end{document}